\documentclass[runningheads]{llncs}
\usepackage[T1]{fontenc}
\usepackage{graphicx}
\usepackage{amsthm,amsmath,amssymb,amsfonts,mathtools,url,soul,verbatim,lineno}

\usepackage{tabulary}
\newcolumntype{C}[1]{>{\centering\arraybackslash}p{#1}}

\theoremstyle{definition}
\newtheorem{dfn}{Definition}

\newtheorem{thm}[dfn]{Theorem}

\newtheorem{cor}[dfn]{Corollary}

\newtheorem{exa}[dfn]{Example}

\newcommand{\R}{\mathbb R}
\newcommand{\Z}{\mathbb Z}

\newcommand{\La}{\Lambda}
\newcommand{\ti}{\tilde}

\newcommand{\vol}{\mathrm{Vol}}
\newcommand{\bs}{\hfill $\blacksquare$}

\newcommand{\vl}{\,:\,}

\begin{document}
\title{Density functions of periodic sequences
}
%
%
\author{
Olga Anosova\inst{1}\orcidID{0000-0003-4134-4398} \and 
Vitaliy Kurlin\inst{1}\orcidID{0000-0001-5328-5351}
}
\authorrunning{O.~Anosova and V.~Kurlin}
%
\institute{Department of Computer Science, University of Liverpool, Liverpool L69 3BX, UK
\email{oanosova@liv.ac.uk, vkurlin@liv.ac.uk}, \url{http://kurlin.org}
}
\maketitle              
\begin{abstract}
Periodic point sets model all solid crystalline materials whose structures are determined in a rigid form and should be studied up to rigid motion or isometry preserving inter-point distances.
In 2021 H.~Edelsbrunner et al. introduced an infinite sequence of density functions that are continuous isometry invariants of periodic point sets.
These density functions turned out to be highly non-trivial even in dimension 1 for periodic sequences of points in the line.
This paper fully describes the density functions of any periodic sequence and their symmetry properties.  
The explicit description theoretically confirms coincidences of density functions that were previously computed only through finite samples.

\end{abstract}

\section{Motivations for density functions of periodic point sets}

Motivated by applications to solid crystalline materials, H.~Edelsbrunner et al.~\cite{edels2021} initiated an isometry classification of periodic point sets.
The most fundamental model of a periodic crystal is a periodic set of points at all atomic centers.
\medskip

Indeed, nuclei of atoms are well-defined physical objects, while chemical bonds are not real sticks or strings but only abstractly represent inter-atomic interactions depending on many thresholds for distances and angles.
\medskip

Since crystal structures are determined in a rigid form, their most practical equivalence is \emph{rigid motion} (a composition of translations and rotations) or \emph{isometry} that maintains all inter-point distances, hence includes reflections \cite{widdowson2022average}.
\medskip

Since atoms always vibrate at any finite temperature above absolute zero, X-ray diffraction patterns of the same material contain inevitable noise and lead to slightly different crystal structures determined at variable temperatures. 
\medskip

In the past, crystallography distinguished periodic structures by coarser isometry invariants such as symmetry groups, which are discontinuous under perturbations \cite[Fig.~1]{widdowson2022average}.
To continuously quantify the similarity between near-duplicates among experimental and simulated structures, we need stronger isometry invariants that continuously change under perturbations \cite[Problem~3]{anosova2021introduction}. 
\medskip

The past work \cite{edels2021} introduced an infinite sequence of density functions $\psi_k[S](t)$ that are continuous isometry invariants of a periodic point set $S$ as defined below.
Let $\R^n$ be the $n$-dimensional Euclidean space, $\Z$ be the set of all integers.

\begin{dfn}[a lattice $\La$, a unit cell $U$, a motif $M$, a periodic set $S=M+\La$]
\label{dfn:crystal}
For a linear basis $ v_1,\dots, v_n$ of $\R^n$, a {\em lattice} is $\La=\{\sum\limits_{i=1}^n c_i v_i : c_i\in\Z\}$.
The \emph{unit cell} $U( v_1,\dots, v_n)=\left\{ \sum\limits_{i=1}^n c_i v_i \vl c_i\in[0,1) \right\}$ is the parallelepiped spanned by the basis.
A \emph{motif} $M\subset U$ is any finite set of points $p_1,\dots,p_m\in U$.
A \emph{periodic point set} \cite{widdowson2022average}
is the Minkowski sum $S=M+\La=\{ u+ v \mid  u\in M,  v\in \La\}$.
\bs
\end{dfn}

In dimension $n=1$, a lattice is defined by any non-zero vector $v\in\R$, any periodic point set $S$ is a periodic sequence $\{p_1,\dots,p_m\}+|v|\Z$ of the period $|v|$.

\begin{dfn}[density functions]
\label{dfn:densities}
Let a periodic set $S=\La+M\subset\R^n$ have a unit cell $U$.
For any integer $k\geq 0$, let $U_k(t)\subset U$ denote the area covered by $k$ closed balls with a radius $t>0$ and centers at all points of $S$.
The $k$-th \emph{density function} is $\psi_k[S](t)=\vol[U_k(t)]/\vol[U]$.
The \emph{density fingerprint} is the sequence $\Psi[S]=\{\psi_k(t)\}_{k=0}^{+\infty}$, see details and examples in \cite[Definition~1 and Figure~2]{edels2021}.
\bs
\end{dfn}

The implementation \cite{edels2021} computes the density functions $\psi_k(t)$ at uniform radii $t$ up to given upper bounds of $t$ and $k$.
This paper explicitly describes all density functions $\psi_k(t)$ for any periodic sequence $S\subset\R$ in Theorems~\ref{thm:0th_density} and~\ref{thm:densities}.
Theorem~\ref{thm:symmetries} proves the symmetry and periodicity of $\psi_k(t)$ in the variables $t$ and $k$.
Corollary~\ref{cor:gen_complete} concludes that the 1st density function $\psi_1(t)$ distinguishes all non-isometric periodic sequences with distinct distances between motif points.

\section{Past work on isometry invariants of periodic point sets}
\label{sec:review}

The strongest result about the density fingerprint $\Psi[S]$ is \cite[Theorem~2]{edels2021} proving that any non-isometric periodic point sets in $\R^3$ have different sequences $\psi_k(t)$, though there was no simple upper bound for $k$.  
However, the density fingerprint turned out to be incomplete \cite[section~5]{edels2021} for the periodic sequences below.

\begin{exa}[periodic sequences $S_{15},Q_{15}\subset\R$]
\label{exa:SQ15}
\cite[Appendix~B]{widdowson2022average} discusses homometric periodic sets that can be distinguished by the recent invariant AMD (Average Minimum Distances) and not by inter-point distance distributions. 
The 
$$\text{sets } 
S_{15} = \{0,1,3,4,5,7,9,10,12\}+15\Z,\;
Q_{15} = \{0,1,3,4,6,8,9,12,14\}+15\Z$$ have the period 15 and the unit cell $[0,15]$ shown as a circle in Fig.~\ref{fig:SQ15}.
\medskip

\begin{figure}[ht]
\includegraphics[width=\textwidth]{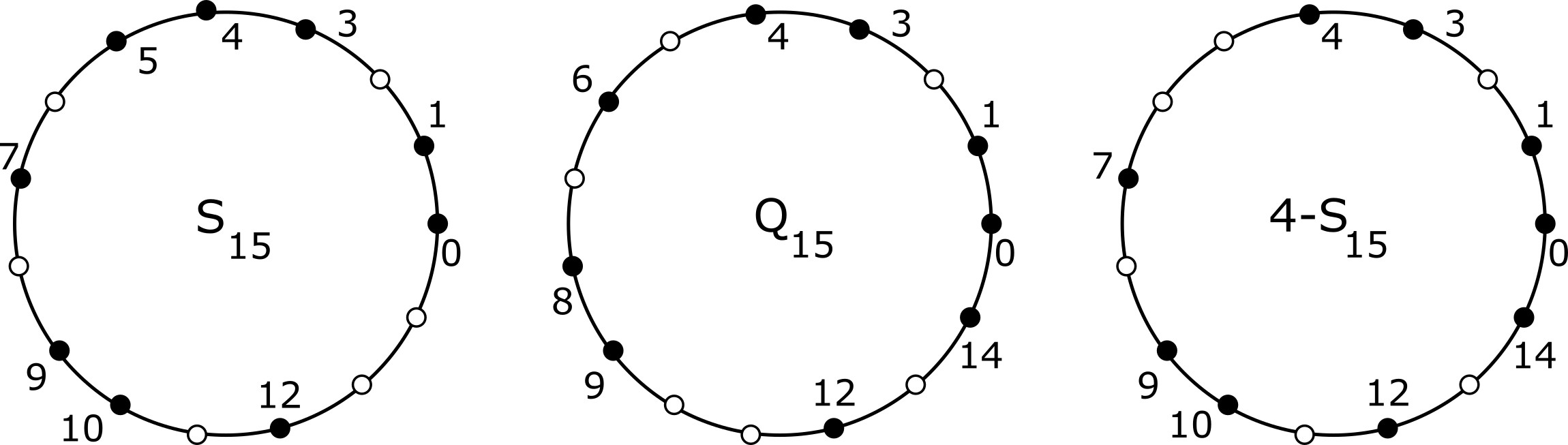}
\caption{Circular versions of the periodic sets $S_{15},Q_{15}$.
Distances are along round arcs.}
\label{fig:SQ15}      
\end{figure}

These periodic sequences \cite{grunbaum1995use} are obtained as Minkowski sums $S_{15}=U+V+15\Z$ and $Q_{15}=U-V+15\Z$ for 
$U = \{0, 4, 9\}$ and $V = \{0, 1, 3\}$.
The last picture in Fig.~\ref{fig:SQ15} shows the periodic set $4-S_{15}$ isometric to $S_{15}$.
Now the difference between $Q_{15}$ and $4-S_{15}$ is better visible: points $0,1,3,4,5,12,14$ are common, but points $6,8,9\in Q_{15}$ are shifted to $7,9,10$ in the circular set $4-S_{15}$.
\bs
\end{exa}

For rational-valued periodic sequences, \cite[Theorem~4]{grunbaum1995use} proved that $r$-th order invariants (combinations of $r$-factor products) up to $r=6$ are enough to distinguish such sequences up to a shift (a rigid motion of $\R$ without reflections).
The AMD invariant was extended to a Pointwise Distance Distribution (PDD), whose generic completeness \cite[Theorem~11]{widdowson2021pointwise} was proved in any dimension $n\geq 1$, but there are finite sets in $\R^3$ with the same PDD \cite[Fig.~S4]{pozdnyakov2020incompleteness}.
In addition to the completeness and continuity under perturbations, 
applications also need a computable metric on isometry classes of periodic point sets.
Such a metric was defined on the complete isoset invariant \cite[section~7]{anosova2021introduction} but has only an approximate algorithm because of a minimization over infinitely many rotations. 
\medskip

This paper fully elucidates all density functions and their exact computation for any periodic sequence, leading to new problems at the end of section~\ref{sec:properties}.

\section{A description of density functions of periodic sequences}
\label{sec:description}

The main results of this section are Theorems~\ref{thm:0th_density} and~\ref{thm:densities} explicitly describing all density functions $\psi_k[S](t)$ for any periodic sequence $S$ of points in $\R$. 
\medskip

\begin{figure}[ht]
\includegraphics[width=\textwidth]{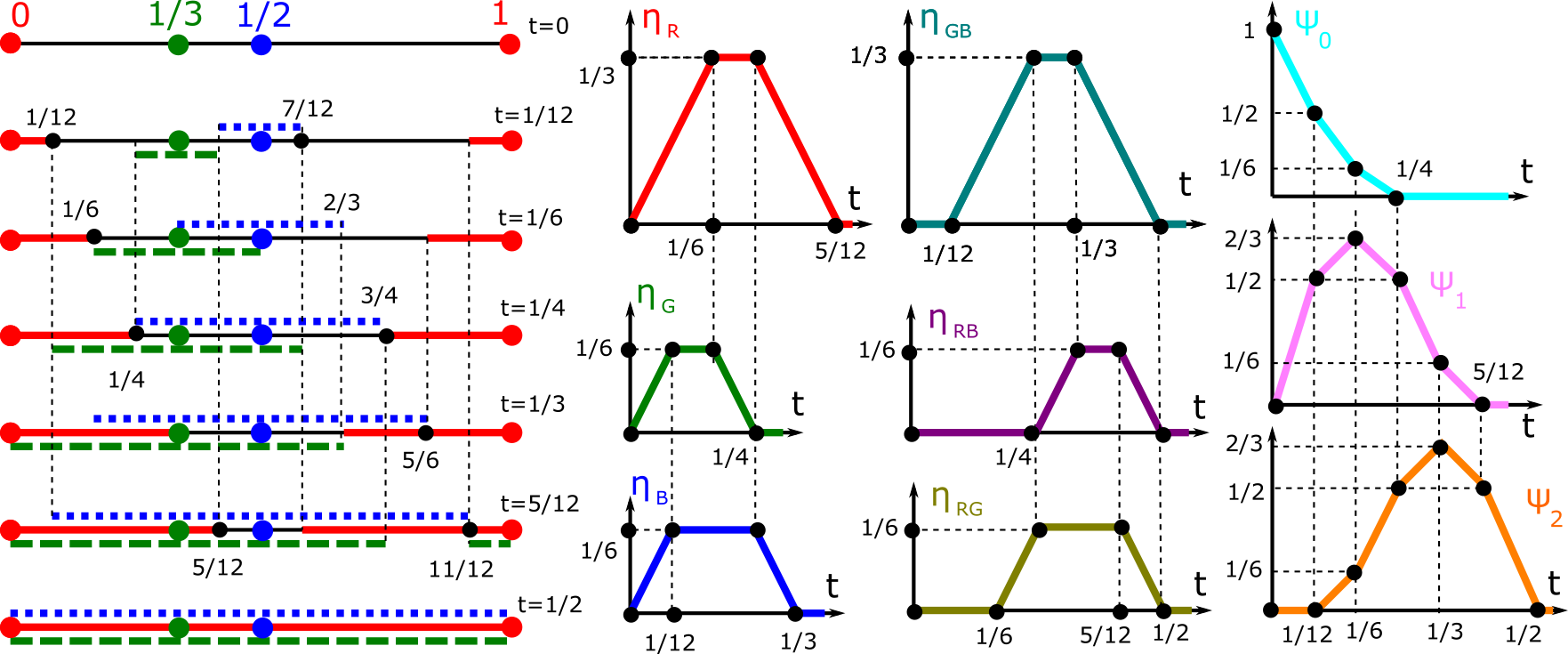}
\caption{\textbf{Left}: the periodic sequence $S=\{0,\frac{1}{3},\frac{1}{2}\}+\Z$ with points of three colors. 
The growing intervals around the red point $0\equiv 1\pmod{1}$, green point $\frac{1}{3}$, blue point $\frac{1}{2}$ have the same color for various radii $t$.
\textbf{Right}: the trapezoid functions $\eta$ from Example~\ref{exa:densities}.
}
\label{fig:densities1D}      
\end{figure}

Since the expanding balls in $\R$ are growing intervals, volumes of their intersections linearly change in the variable radius $t$.
Hence any density function $\psi_k(t)$ is piecewise linear and uniquely determined by \emph{corner} points $(a_j,b_j)$ where the gradient changes.
Examples~\ref{exa:0th_density} and~\ref{exa:densities} explain
how the density functions $\psi_k(t)$ are computed for the periodic sequence $S=\{0,\frac{1}{3},\frac{1}{2}\}+\Z$, see all graphs in Fig.~\ref{fig:densities1D}.

\begin{exa}[$0$-th density $\psi_0(t)$ for $S=\{0,\frac{1}{3},\frac{1}{2}\}+\Z$]
\label{exa:0th_density}
By Definition~\ref{dfn:densities} $\psi_0(t)$ is the fractional length within the period interval $[0,1]$ not covered by the intervals of radius $t$ (length $2t$), which are the red intervals $[0,t]\cup[1-t,1]$, green dashed interval $[\frac{1}{3}-t,\frac{1}{3}+t]$ and blue dotted interval $[\frac{1}{2}-t,\frac{1}{2}+t]$.
The graph of $\psi_0(t)$ starts from the point $(0,1)$ at $t=0$.
Then $\psi_0(t)$ linearly drops to the point $(\frac{1}{12},\frac{1}{3})$ at $t=\frac{1}{12}$ when a half of the interval $[0,1]$ remains uncovered.
\medskip

The next linear piece of $\psi_0(t)$ continues to the point $(\frac{1}{6},\frac{1}{6})$ at $t=\frac{1}{6}$ when only $[\frac{2}{3},\frac{5}{6}]$ is uncovered.
The graph of $\psi_0(t)$ finally returns to the $t$-axis at the point $(\frac{1}{4},0)$ and remains there for $t\geq \frac{1}{4}$.
The piecewise linear behavior of $\psi_0(t)$ can be briefly described via the \emph{corner} points $(0,1)$, $(\frac{1}{12},\frac{1}{3})$, $(\frac{1}{6},\frac{1}{6})$, $(\frac{1}{4},0)$.
\bs
\end{exa}

Theorem~\ref{thm:0th_density} extends Example~\ref{exa:0th_density}
 to any periodic sequence $S$ and implies that $\psi_0(t)$ is uniquely determined by the ordered distances within a unit cell of $S$.

\vspace*{-6mm}
\begin{thm}[description of $\psi_0$]
\label{thm:0th_density}
Any periodic sequence $S\subset\R$ can be scaled to period 1 so that $S=\{p_1,\dots,p_m\}+\Z$.
Set $d_i=p_{i+1}-p_i\in(0,1)$, where $i=1,\dots,m$ and $p_{m+1}=p_1+1$. 
Put the distances in the increasing order $d_{[1]}\leq d_{[2]}\leq\dots\leq d_{[m]}$.
Then the 0-th density function $\psi_0$ is piecewise linear with the following (unordered) corners: $(0,1)$ and $(\frac{1}{2}d_{[i]}, 1-\sum\limits_{j=1}^{i-1} d_{[j]}-(m-i+1)d_{[i]})$ for $i=1,\dots,m$, so the last corner is $(\frac{1}{2}d_{[m]},0)$.
If any corner points are repeated, e.g. when $d_{[i-1]}=d_{[i]}$, these corners are collapsed into one corner point. 
\bs
\end{thm}
\begin{proof}
The function $\psi_0(t)$ measures the total length of subintervals in $[0,1]$ that are not covered by growing intervals $[p_i-t,p_i+t]$, $i=1,\dots,m$. 
Hence $\psi_0(t)$ linearly decreases from the initial value $\psi_0(0)=1$ except for $m$ critical values of $t$ where one of the intervals $[p_i,p_{i+1}]$ between successive points become completely covered and can not longer shrink.
These critical radii $t$ are ordered according to the distances $d_{[1]}\leq d_{[2]}\leq\dots\leq d_{[m]}$.
The first critical radius is $t=\frac{1}{2}d_{[1]}$, when the shortest interval $[p_i,p_{i+1}]$ of the length $d_{[1]}$ is covered by the intervals centered at $p_i,p_{i+1}$.
At this moment, all $m$ intervals cover the subregion of the length $md_{[1]}$.
Then $\psi_0(t)$ has the first corner points $(0,1)$ and $(\frac{1}{2}d_{[1]},1-md_{[1]})$. 

The second critical radius is $t=\frac{1}{2}d_{[2]}$, when the covered subregion has the length $d_{[1]}+(m-1)d_{[2]}$, i.e. the next corner point is $(\frac{1}{2}d_{[2]},1-d_{[1]}-(m-1)d_{[2]})$. 
If $d_{[1]}=d_{[2]}$, then both corner points coincide, so $\psi_0(t)$ will continue from the joint corner point.
The above pattern generalizes to the $i$-th critical radius $t=\frac{1}{2}d_{[i]}$, when the covered subregion has the length $\sum\limits_{j=1}^{i-1}d_{[j]}$ (for the finally covered intervals) plus $(m-i+1)d_{[i]}$ (for the still growing intervals).
For the final critical radius $t=\frac{1}{2}d_{[m]}$, the whole interval $[0,1]$ is covered by the grown intervals because $\sum\limits_{j=1}^{m}d_{[j]}=1$.
So the final corner point of $\psi_0(t)$ is $(\frac{1}{2}d_{[m]},0)$.
\end{proof}

Theorem~\ref{thm:0th_density} for the sequence $S=\{0,\frac{1}{3},\frac{1}{2}\}+\Z$ gives the ordered distances $d_{[1]}=\frac{1}{6}<d_{[2]}=\frac{1}{3}<d_{[3]}=\frac{1}{2}$, which determine the corner points $(0,1)$, $(\frac{1}{12},\frac{1}{2})$, $(\frac{1}{6},\frac{1}{6})$, $(\frac{1}{4},0)$ of the density function $\psi_0(t)$ in Fig.~\ref{fig:densities1D}, see Example~\ref{exa:0th_density}.
\medskip

By Theorem~\ref{thm:0th_density} any 0th density function $\psi_0(t)$ is uniquely determined by the (unordered) set of lengths of intervals between successive points.
Hence we can re-order these intervals without changing $\psi_0(t)$.
For instance, the periodic sequence $Q=\{0,\frac{1}{2},\frac{2}{3}\}+\Z$ has the same set of interval lengths $d_{[1]}=\frac{1}{6}$, $d_{[2]}=\frac{1}{3}$, $d_{[3]}=\frac{1}{2}$ as the periodic sequence $S=\{0,\frac{1}{3},\frac{1}{2}\}+\Z$ in Example~\ref{exa:0th_density}.
\medskip

The above sequences $S,Q$ are related by the mirror reflection $t\mapsto 1-t$.
One can easily construct many non-isometric sequences with $\psi_0[S](t)=\psi_0[Q](t)$.
For any $1\leq i\leq m-3$, the sequences $S_{m,i}=\{0,2,3,\dots,i+2,i+4,i+5,\dots,m+2\}+(m+2)\Z$ have the same interval lengths $d_{[1]}=\dots=d_{[m-2]}=1$, $d_{[m-1]}=d_{[m]}=2$ but are not related by isometry (translations and reflections in $\R$) because the intervals of length 2 are separated by $i-1$ intervals of length 1 in $S_{m,i}$.  
\medskip

Corollary~\ref{cor:gen_complete} will prove that the 1st density function $\psi_1[S](t)$  uniquely determines a periodic sequence $S\subset\R$ in general position up to isometry of $\R$.

\begin{exa}[functions $\psi_k(t)$ for $S=\{0,\frac{1}{3},\frac{1}{2}\}+\Z$]
\label{exa:densities}
The 1st density function $\psi_1(t)$ can be obtained as a sum of the three \emph{trapezoid} functions $\eta_R$, $\eta_G$, $\eta_B$, each measuring the length of a region covered by a single interval (of one color).
The red intervals $[0,t]\cup[1-t,1]$ grow until $t=\frac{1}{6}$ when they touch the green interval $[\frac{1}{6},\frac{1}{2}]$.
So the length $\eta_R(t)$ of this interval linearly grows from the origin $(0,0)$ to the corner point $(\frac{1}{6},\frac{1}{3})$.
For $t\in[\frac{1}{6},\frac{1}{4}]$, the left red interval is shrinking at the same rate due to the overlapping green interval, while the right red interval continues to grow until $t=\frac{1}{4}$, when it touches the blue interval $[\frac{1}{4},\frac{3}{4}]$.  
Hence the graph of $\eta_R(t)$ remains constant up to the corner point $(\frac{1}{4},\frac{1}{3})$.
After that $\eta_R(t)$ linearly returns to the $t$-axis at $t=\frac{5}{12}$.
Hence the trapezoid function $\eta_R$ has the piecewise linear graph through the corner points $(0,0)$, $(\frac{1}{6},\frac{1}{3})$, $(\frac{1}{4},\frac{1}{3})$, $(\frac{5}{12},\frac{1}{0})$. 
\medskip

The 2nd function $\psi_2(t)$ is the sum of the \emph{trapezoid} functions $\eta_{GB},\eta_{RG},\eta_{RB}$, each measuring the length of a double intersection.
For the green interval $[\frac{1}{3}-t,\frac{1}{3}+t]$ and the blue interval $[\frac{1}{2}-t,\frac{1}{2}+t]$, the graph of the trapezoid function $\eta_{GB}(t)$ is piecewise linear and starts at the point $(\frac{1}{12},0)$, where the intervals touch.
The green-blue intersection interval $[\frac{1}{2}-t,\frac{1}{3}+t]$ grows until $t=\frac{1}{4}$, when $[\frac{1}{4},\frac{7}{12}]$ touches the red interval on the left.
At the same time $\eta_{GB}(t)$ is linearly growing to the point $(\frac{1}{4},\frac{1}{3})$.
For $t\in[\frac{1}{4},\frac{1}{3}]$, the green-blue intersection interval becomes shorter on the left, but grows at the same rate on the right until $[\frac{1}{3},\frac{2}{3}]$ touches the red interval $[\frac{2}{3},1]$.
Then $\eta_{GB}(t)$ remains constant up to the point $(\frac{1}{3},\frac{1}{3})$.
For $t\in[\frac{1}{3},\frac{1}{2}]$ the green-blue intersection interval is shortening from both sides.
Finally, the graph of $\eta_{GB}(t)$ returns to the $t$-axis at $(\frac{1}{2},0)$, see Fig.~\ref{fig:densities1D}.
\bs
\end{exa}

Theorem~\ref{thm:densities} extends Example~\ref{exa:densities} 
and proves that any $\psi_k(t)$ is a sum of trapezoid functions whose corners are explicitly described. 
We consider any index $i=1,\dots,m$ (of a point $p_i$ or a distance $d_i$) modulo $m$ so that $m+1\equiv 1\pmod{m}$.

\begin{thm}[description of $\psi_k$, $k>0$]
\label{thm:densities}
Since any periodic point set $S\subset\R$ can be scaled to unit cell $[0,1]$, let $S=\{p_1,\dots,p_m\}+\Z$ and $d_i=p_{i+1}-p_i\in(0,1)$, where $i=1,\dots,m$ and $p_{m+1}=p_1+1$. 
Any interval $[p_i-t,p_i+t]$ is projected to $[0,1]$ modulo $\Z$.
For $1\leq k\leq m$, the density function $\psi_k(t)$ is the sum of $m$ \emph{trapezoid} functions $\eta_{k,i}$ with the corner points 
$(\frac{s}{2},0)$, 
$(\frac{d_{i-1}+s}{2},d)$, 
$(\frac{s+d_{i+k-1}}{2},d)$, 
$(\frac{d_{i-1}+s+d_{i+k-1}}{2},0)$, where 
$d=\min\{d_{i-1},d_{i+k-1}\}$, $s=\sum\limits_{j=i}^{i+k-2}d_j$,
$i=2,\dots,m+1$.
If $k=1$, then $s=0$ is the empty sum.
So $\psi_k(t)$ is determined by the unordered set of triples $(d_{i-1},s,d_{i+k-1})$ whose first and last entries are swappable.
\bs 
\end{thm}
\begin{proof}
For simplicity, we separately prove the case $k=1$.
The 1st density function $\psi_1(t)$ measures the total length of subregions covered by a single interval $[p_i-t,p_i+t]$.
Hence $\psi_1(t)$ is the sum of the functions $\eta_{1i}$, each measuring the length of the subinterval of $[p_i-t,p_i+t]$ not covered by other such intervals.
\medskip

Each function $\eta_{1i}$ starts from $\eta_{1i}(0)=0$ and linearly grows up to $\eta_{1i}(\frac{1}{2}d)=d$, where $d=\min\{d_{i-1},d_{i}\}$, when the interval $[p_i-t,p_i+t]$ of the length $2t=d$ touches the growing interval centered at the closest of its neighbors $p_{i\pm 1}$.
\medskip

If (say) $d_{i-1}<d_i$, then the subinterval covered only by $[p_i-t,p_i+t]$ is shrinking on the left and is growing at the same rate on the right until it touches the growing interval centered at the right neighbor.
During this period, when $t$ is between $\frac{1}{2}d_{i-1}$ and $\frac{1}{2}d_i$, the trapezoid function $\eta_{1i}(t)=d$ remains constant.
\medskip

If $d_{i-1}=d_i$, this horizontal piece collapses to one point in the graph of $\eta_{1i}(t)$.
For $t\geq\max\{d_{i-1},d_{i}\}$, the subinterval covered only by $[p_i-t,p_i+t]$ is shrinking on both sides until the intervals centered at $p_{i\pm 1}$ meet at a mid-point between them for $t=\frac{d_{i-1}+d_i}{2}$. 
So the graph of $\eta_{1i}$ has a trapezoid form with the corner points $(0,0)$, $(\frac{d_{i-1}}{2}, d)$, $(\frac{d_{i}}{2},d)$, $(\frac{d_{i-1}+d_{i}}{2},0)$.
\medskip

In Example~\ref{exa:densities} for $S=\{0,\frac{1}{3},\frac{1}{2}\}+\Z$, the distances $d_{1}=\frac{1}{3}$, $d_{2}=\frac{1}{6}$, $d_{3}=\frac{1}{2}=d_0$ give $\eta_{11}=\eta_{R}$ with the corner points $(0,0)$, $(\frac{1}{4},\frac{1}{3})$, $(\frac{1}{6},\frac{1}{3})$, $(\frac{5}{12},0)$ as in Fig.~\ref{fig:densities1D}.
\bigskip

\noindent
In the case $k>1$, the $k$-th density function $\psi_k(t)$ measures the total length of $k$-fold intersections among $m$ intervals $[p_i-t,p_i+t]$, $i=1,\dots,m$.
\medskip

A $k$-fold intersection appears only when two intervals $[p_i-t,p_i+t]$ and $[p_{i+k-1}-t,p_{i+k-1}+t]$ overlap because their intersection is covered by the $k$ intervals centered at $k$ points $p_i<p_{i+1}<\dots<p_{i+k-1}$.
Since only $k$ successive intervals can contribute to $k$-fold intersections, $\psi_k(t)$ becomes the sum of the functions $\eta_{k,i}$, each equal to the length of the subinterval of $[p_i-t,p_{i+k-1}+t]$ covered by exactly $k$ intervals of the form $[p_j-t,p_j+t]$, $j=1,\dots,m$.
\medskip

The above function $\eta_{k,i}(t)$ remains 0 until the radius $t=\frac{1}{2}\sum\limits_{j=i}^{i+k-2}d_j$ because $2t$ is the length between the points $p_i<p_{i+k-1}$.
Then $\eta_{k,i}(t)$ is linearly growing until the $k$-fold intersection touches one of the intervals centered at the points $p_{i-1},p_{i+k}$, which are left and right neighbors of $p_i,p_{i+k-1}$, respectively.
\medskip

If (say) $d_{i-1}<d_{i+k-1}$, this critical radius is $t=\frac{1}{2}\sum\limits_{j=i-1}^{i+k-2}d_j=\frac{d_{i-1}+s}{2}$.
The function $\eta_{k,i}(t)$ measures the length of the $k$-fold intersection $[p_{i+k-1}-t,p_i+t]$.
$$\eta_{k,i}(t)=(p_i+t)-(p_{i+k-1}-t)=2t-(p_{i+k-1}-p_i)=(d_{i-1}+s)-s=d_{i-1}.$$
Then the $k$-fold intersection is shrinking on the left and is growing at the same rate on the right until it touches the growing interval centered at the right neighbor $p_{i+k}$.
During this time, when $t$ is between $\frac{1}{2}\sum\limits_{j=i-1}^{i+k-2}d_j$ and $\frac{1}{2}\sum\limits_{j=i}^{i+k-1}d_j$, the function $\eta_{k,i}(t)$ remains equal to $d_{i-1}$.
If $d_{i-1}>d_{i+k-1}$, the last argument should include the smaller distance $d_{i+k-1}$ instead of $d_{i-1}$.
Hence we will use below the single value $d=\min\{d_{i-1},d_{i+k-1}\}$ to cover both cases. 
If $d_{i-1}=d_i$, this horizontal piece collapses to one point in the graph of $\eta_{k,i}(t)$.
The $k$-fold intersection within $[p_i,p_{i+k-1}]$ disappears when the intervals centered at $p_{i-1},p_{i+k}$ have the radius $t$ equal to the half-distance $\frac{1}{2}\sum\limits_{j=i-1}^{i+k-1}d_j$ between $p_{i-1},p_{i+k}$.
\medskip

Then $\eta_{k,i}(t)$ is the trapezoid function with the expected four corner points expressed as $(\frac{s}{2},0)$, $(\frac{d_{i-1}+s}{2},d)$, $(\frac{s+d_{i+k-1}}{2},d)$, $(\frac{d_{i-1}+s+d_{i+k-1}}{2},0)$ for $s=\sum\limits_{j=i}^{i+k-2}d_j$ and $d=\min\{d_{i-1},d_{i+k-1}\}$.
These corners are uniquely determined by the triple $(d_{i-1},s,d_{i+k-1})$, where the components $d_{i-1},d_{i+k-1}$ can be swapped. 
\end{proof}

In Example~\ref{exa:densities} for $S=\{0,\frac{1}{3},\frac{1}{2}\}+\Z$, we have $d_{1}=\frac{1}{3}$, $d_{2}=\frac{1}{6}$, $d_{3}=\frac{1}{2}=d_0$.
For $k=2$, $i=2$, we get $d_{i-1}=d_1=\frac{1}{3}$, $d_{i+k-1}=d_3=\frac{1}{2}$, i.e. $d=\min\{d_1,d_3\}=\frac{1}{3}$, $s=d_{2}=\frac{1}{6}$.
Then $\eta_{22}=\eta_{GB}$ has the corner points $(\frac{1}{12},0)$, $(\frac{1}{4},\frac{1}{3})$, $(\frac{1}{3},\frac{1}{3})$, $(\frac{1}{2},0)$. 

\section{Symmetries, computations, and generic completeness}
\label{sec:properties}

\begin{thm}[symmetries of $\psi_k(t)$]
\label{thm:symmetries}
For any periodic sequence $S\subset\R$ with a unit cell $[0,1]$, we have the \emph{periodicity} 
$\psi_{k+m}(t+\frac{1}{2})=\psi_{k}(t)$ for any $k\geq 0$, $t\geq 0$, and the \emph{symmetry} $\psi_{m-k}(\frac{1}{2}-t)=\psi_k(t)$ for $k=0,\dots,[\frac{m}{2}]$, and $t\in[0,\frac{1}{2}]$.
\bs
\end{thm}
\begin{proof}
To prove $\psi_{m-k}(\frac{1}{2}-t)=\psi_k(t)$ for $k=1,\dots,[\frac{m}{2}]$, we establish a bijection between the triples of parameters that determined $\psi_{m-k}$ and $\psi_k$ in Theorem~\ref{thm:densities}.
\medskip

Take a triple $(d_{i-1},s,d_{i+k-1})$ of $\psi_k$, where $s=\sum\limits_{j=i}^{i+k-2}d_j$ is the sum of $k-1$ distances from $d_{i-1}$ to $d_{i+k-1}$ in the increasing (cyclic) order of distance indices.
Under $t\mapsto\frac{1}{2}-t$, the corner points of trapezoid function $\eta_{k,i}$ map to
$$\Big(\frac{1-s}{2},0\Big),\;
\Big(\frac{1-s-d_{i-1}}{2},d\Big),\; 
\Big(\frac{1-s-d_{i+k-1}}{2},d\Big),\;
\Big(\frac{1-d_{i-1}-s-d_{i+k-1}}{2},0\Big).$$

Notice that $\bar s=1-d_{i-1}-s-d_{i+k-1}$ is the sum of the $m-k-1$ intermediate distances from $d_{i+k-1}$ to $d_{i-1}$ in the increasing (cyclic) order of indices.
\medskip

The four corner points can be re-written with the above notation $\bar s$ as follows:
$$\left(\frac{d_{i-1}+\bar s+d_{i+k-1}}{2},0\right),\quad 
\left(\frac{\bar s+d_{i+k-1}}{2},d\right),\quad
\left(\frac{\bar s+d_{i-1}}{2},d\right),\quad
\left(\frac{\bar s}{2},0\right).$$
These resulting points are re-ordered corners of the trapezoid function $\eta_{m-k,i+k}$.
Hence $\eta_{k,i}(\frac{1}{2}-t)=\eta_{m-k,i+k}(t)$.
Taking the sum over all indices $i=1,\dots,m$, we get $\psi_{k}(\frac{1}{2}-t)=\psi_{m-k}(t)$.
Fig.~\ref{fig:densities1D} shows the symmetry  
$\psi_1(t)=\psi_2(\frac{1}{2}-t)$.
\bigskip

For periodicity, we compare $\psi_k$ and $\psi_{k+m}$ for $k\geq 0$.
Any $(k+m)$-fold intersection should involve intervals centered at $k+m$ successive points of the sequence $S\subset\R$.
Then we can find a period interval $[t,t+1]$ covering $m$ of these points.
By collapsing this interval to a single point, the $(k+m)$-fold intersection becomes $k$-fold, but its fractional length within any period interval of length 1 remains the same.
Since the radius $t$ is twice smaller than the length of the corresponding interval, this collapse gives us
$\psi_{k+m}(t+\frac{1}{2})=\psi_{k}(t)$.
\smallskip

The final symmetry $\psi_{m}(\frac{1}{2}-t)=\psi_0(t)$ follows from $\psi_{m}(\frac{1}{2}-t)=\psi_{m}(\frac{1}{2}+t)$.
Indeed, any trapezoid of $\psi_{m}$ has $s=1-d_{i-1}$.
Since its four corners $(\frac{1-d_{i-1}}{2},0)$, $(\frac{1}{2},\frac{d_{i-1}}{2})$, $(\frac{1}{2},\frac{d_{i-1}}{2})$, $(\frac{1+d_{i-1}}{2},0)$ are symmetric in $t=\frac{1}{2}$, then so is the sum $\psi_m$. 
\end{proof}

\begin{cor}[computation of $\psi_k(t)$]
\label{cor:computation}
Let $S,Q\subset\R$ be periodic sequences with at most $m$ motif points.
For $k\geq 1$, one can draw the graph of the $k$-th density function $\psi_k[S]$ in time $O(m^2)$.
One can check in time $O(m^3)$ if 
$\Psi[S]=\Psi[Q]$.
\bs
\end{cor}
\begin{proof}
To draw the graph of $\psi_k[S]$ or evaluate the $k$-th density function $\psi_k[S](t)$ at any $t$, we first use the symmetry and periodicity from Theorem~\ref{thm:symmetries} to reduce $k$ to the range $0,1,\dots,[\frac{m}{2}]$.
In time $O(m\log m)$ we put the points from a unit cell $U$ (scaled to $[0,1]$ for convenience) in the increasing (cyclic) order $p_1,\dots,p_m$.
In time $O(m)$ we compute the distances $d_i=p_{i+1}-p$ between successive points.
\smallskip

For $k=0$, we put the distances in the increasing order $d_{[1]}\leq\dots\leq d_{[m]}$ in time $O(m\log m)$.
By Theorem~\ref{thm:0th_density} in time $O(m^2)$, we write down the $O(m)$ corner points whose horizontal coordinates are the critical radii where $\psi_0(t)$ can change its gradient. 
We evaluate $\psi_0$ at every critical radius $t$ by summing up the values of $m$ trapezoid functions at $t$, which needs $O(m^2)$ time.
It remains to plot the points at all $O(m)$ critical radii $t$ and connect the successive points by straight lines, so the total time is $O(m^2)$.
For any larger fixed index $k=1,\dots,[\frac{m}{2}]$, in time $O(m^2)$ we write down all $O(m)$ corner points from Theorem~\ref{thm:densities}, which leads to the graph of $\psi_k(t)$ similarly to the above argument for $k=0$.
\smallskip

To decide if the infinite sequences of density functions coincide: $\Psi[S]=\Psi[Q]$, by Theorem~\ref{thm:symmetries} it suffices to check only if $O(m)$ density functions coincide: $\psi_k[S](t)=\psi_k[Q](t)$ for $k=0,1,\dots,[\frac{m}{2}]$.
To check if two piecewise linear functions coincide, it remains to compare their values at all $O(m)$ critical radii $t$ from the corner points in Theorems~\ref{thm:0th_density} and \ref{thm:densities}.
Since these values were found in time $O(m^2)$ above, the total time for $k=0,1,\dots,[\frac{m}{2}]$ is $O(m^3)$. 
\end{proof}

To illustrate Corollary~\ref{cor:computation}, Example~\ref{exa:SQ15densities} will justify that the periodic sequences $S_{15}$ and $Q_{15}$ in Fig.~\ref{fig:SQ15} have identical density fingerprints $\Psi[S_{15}]=\Psi[Q_{15}]$.

\begin{exa}[$S_{15},Q_{15}$ have equal density functions]
\label{exa:SQ15densities}
To avoid fractions, we keep the unit cell $[0,15]$ of the sequences $S_{15},Q_{15}$ 
because all quantities in Theorem~\ref{thm:densities} can be scaled up by factor 15. 
To conclude that $\psi_0[S_{15}]=\psi_0[Q_{15}]$, by 
Theorem~\ref{thm:0th_density} we check that $S_{15},Q_{15}$ have the same set of the ordered distances $d_{[i]}$ between successive points, which is shown in identical rows 3 of Tables~\ref{tab:S15d} and~\ref{tab:Q15d}.

\vspace*{-5mm}
\begin{table}[h!]
\caption{\textbf{Row 1}: points $p_i$ from the set $S_{15}$ in Fig.~\ref{fig:SQ15}.
\textbf{Row 2}: the distances $d_i$ between successive points of $S_{15}$. 
\textbf{Row 3}: the distances $d_{[i]}$ are in the increasing order.
\textbf{Row 4}: the unordered set of these pairs determines the density function $\psi_1$ by Theorem~\ref{thm:densities}. 
\textbf{Row 5}: the pairs are lexicographically ordered for comparison with row 5 in Table~\ref{tab:Q15d}. 
\textbf{Rows 6,8,10}: the unordered sets of these triples determine the density functions $\psi_2,\psi_3,\psi_4$ by Theorem~\ref{thm:densities} for $k=2,3,4$.
\textbf{Rows 7,9,11}: the triples from rows 6,8,10 are ordered for easier comparison with corresponding rows 7,9,11 in Table~\ref{tab:Q15d}, see  details in Example~\ref{exa:SQ15densities}.
}
\label{tab:S15d}
\begin{tabular}{C{30mm}|C{9mm}C{9mm}C{9mm}C{9mm}C{9mm}C{9mm}C{9mm}C{9mm}C{9mm}}
$p_i$                      & 0 & 1 & 3 & 4 & 5 & 7 & 9 & 10 & 12 \\
\hline
\hline
$d_i=p_{i+1}-p_i$ & 1 & 2 & 1 & 1 & 2 & 2 & 1 &  2 & 3 \\
ordered $d_{[i]}$ & 1 & 1 & 1 & 1 & 2 & 2 & 2 &  2 & 3 \\
\hline
\hline
$(d_{i-1},d_i)$  & (3,1) & (1,2) & (2,1) & (1,1) & (1,2) & (2,2) & (2,1) & (1,2) & (2,3) \\
order $(d_{i-1},d_i)$  & (1,1) & (1,2) & (1,2) & (1,2) & (1,2) & (1,2) & (1,3) & (2,2) & (2,3) \\
\hline
\hline
$(d_{i-1},\mathbf{d_i},d_{i+1})$  & (3,{\bf 1},2) & (1,{\bf 2},1) & (2,{\bf 1},1) & (1,{\bf 1},2) & (1,{\bf 2},2) & (2,{\bf 2},1) & (2,{\bf 1},2) & (1,{\bf 2},3) & (2,{\bf 3},1) \\
order $(d_{i-1},\mathbf{d_i},d_{i+1})$ & (1,{\bf 1},2)  & (1,{\bf 1},2) & (2,{\bf 1},2) & (2,{\bf 1},3)  & (1,{\bf 2},1) & (1,{\bf 2},2) & (1,{\bf 2},2) & (1,{\bf 2},3) & (1,{\bf 3},2) \\
\hline
\hline
$(d_{i-1},\mathbf{s},d_{i+2})$  & (3,{\bf 3},1) & (1,{\bf 3},1) & (2,{\bf 2},2) & (1,{\bf 3},2) & (1,{\bf 4},1) & (2,{\bf 3},2) & (2,{\bf 3},3) & (1,{\bf 5},1) & (2,{\bf 4},2) \\
order $(d_{i-1},\mathbf{s},d_{i+2})$ & (2,{\bf 2},2) & (1,{\bf 3},1) & (1,{\bf 3},2) & (1,{\bf 3},3)  & (2,{\bf 3},2)  & (2,{\bf 3},3) & (1,{\bf 4},1) & (2,{\bf 4},2) & (1,{\bf 5},1) \\
\hline
\hline
$(d_{i-1},\mathbf{s},d_{i+3})$  & (3,{\bf 4},1) & (1,{\bf 4},2) & (2,{\bf 4},2) & (1,{\bf 5},1) & (1,{\bf 5},2) & (2,{\bf 5},3) & (2,{\bf 6},1) & (1,{\bf 6},2) & (2,{\bf 6},1) \\
order $(d_{i-1},\mathbf{s},d_{i+3})$  & (1,{\bf 4},2) & (1,{\bf 4},3) & (2,{\bf 4},2) & (1,{\bf 5},1) & (1,{\bf 5},2) & (2,{\bf 5},3) & (1,{\bf 6},2) & (1,{\bf 6},2) & (1,{\bf 6},2) 
\end{tabular}
\end{table}

\vspace*{-5mm}
\begin{table}[h!]
\caption{\textbf{Row 1}: points $p_i$ from the set $Q_{15}$ in Fig.~\ref{fig:SQ15}.
\textbf{Row 2}: the distances $d_i$ between successive points of $Q_{15}$. 
\textbf{Row 3}: the distances $d_{[i]}$ are in the increasing order.
\textbf{Row 4}: the unordered set of these pairs determines the density function $\psi_1$ by Theorem~\ref{thm:densities}b. 
\textbf{Row 5}: the pairs are lexicographically ordered for comparison with row 5 in Table~\ref{tab:S15d}. 
\textbf{Rows 6,8,10}: the unordered sets of these triples determine the density functions $\psi_2,\psi_3,\psi_4$ by Theorem~\ref{thm:densities} for $k=2,3,4$.
\textbf{Rows 7,9,11}: the triples from rows 6,8,10 are ordered for comparison with corresponding rows 7,9,11 in Table~\ref{tab:S15d}, see  Example~\ref{exa:SQ15densities}.
}
\label{tab:Q15d}
\begin{tabular}{C{30mm}|C{9mm}C{9mm}C{9mm}C{9mm}C{9mm}C{9mm}C{9mm}C{9mm}C{9mm}}
\hline
$p_i$                      & 0 & 1 & 3 & 4 & 6 & 8 & 9 & 12 & 14 \\
\hline
\hline
$d_i=p_{i+1}-p_i$ & 1 & 2 & 1 & 2 & 2 & 1 & 3 &  2 & 1 \\
ordered $d_{[i]}$ & 1 & 1 & 1 & 1 & 2 & 2 & 2 &  2 & 3 \\
\hline
\hline
$(d_{i-1},d_i)$  & (1,1) & (1,2) & (2,1) & (1,2) & (2,2) & (2,1) & (1,3) & (3,2) & (2,1) \\
ordered $(d_{i-1},d_i)$  & (1,1) & (1,2) & (1,2) & (1,2) & (1,2) & (1,2) & (1,3) & (2,2) & (2,3) \\
\hline
\hline
$(d_{i-1},\mathbf{d_i},d_{i+1})$  & (1,{\bf 1},2) & (1,{\bf 2},1) & (2,{\bf 1},2) & (1,{\bf 2},2) & (2,{\bf 2},1) & (2,{\bf 1},3) & (1,{\bf 3},2) & (3,{\bf 2},1) & (2,{\bf 1},1) \\
order $(d_{i-1},\mathbf{d_i},d_{i+1})$ & (1,{\bf 1},2)  & (1,{\bf 1},2) & (2,{\bf 1},2) & (2,{\bf 1},3)  & (1,{\bf 2},1) & (1,{\bf 2},2) & (1,{\bf 2},2) & (1,{\bf 2},3) & (1,{\bf 3},2) \\
\hline
\hline
$(d_{i-1},\mathbf{s},d_{i+2})$  & (1,{\bf 3},1) & (1,{\bf 3},2) & (2,{\bf 3},2) & (1,{\bf 4},1) & (2,{\bf 3},3) & (2,{\bf 4},2) & (1,{\bf 5},1) & (3,{\bf 3},1) & (2,{\bf 2},2) \\
order $(d_{i-1},\mathbf{s},d_{i+2})$ & (2,{\bf 2},2) & (1,{\bf 3},1) & (1,{\bf 3},2) & (1,{\bf 3},3)  & (2,{\bf 3},2)  & (2,{\bf 3},3) & (1,{\bf 4},1) & (2,{\bf 4},2) & (1,{\bf 5},1) \\
\hline
\hline
$(d_{i-1},\mathbf{s},d_{i+3})$  & (1,{\bf 4},2) & (1,{\bf 5},2) & (2,{\bf 5},1) & (1,{\bf 5},3) & (2,{\bf 6},2) & (2,{\bf 6},1) & (1,{\bf 6},1) & (3,{\bf 4},2) & (2,{\bf 4},1) \\
order $(d_{i-1},\mathbf{s},d_{i+3})$  & (1,{\bf 4},2) & (1,{\bf 4},2) & (2,{\bf 4},3) & (1,{\bf 5},2) & (1,{\bf 5},2) & (1,{\bf 5},3)  & (1,{\bf 6},1) & (1,{\bf 6},2) & (2,{\bf 6},2)
\end{tabular}
\end{table} 

To conclude that $\psi_1[S_{15}]=\psi_1[Q_{15}]$ by Theorem~\ref{thm:densities}, we check that $S_{15},Q_{15}$ have the same set of unordered pairs $(d_{i-1},d_i)$ of distances between successive points.
Indeed, Tables~\ref{tab:S15d} and~\ref{tab:Q15d} have identical rows 5, where pairs are \emph{lexicograpically} ordered for comparison: $(a,b)<(c,d)$ if $a<b$ or $a=b$ and $c<d$.
\medskip

To conclude that $\psi_k[S_{15}]=\psi_k[Q_{15}]$ for $k=2,3,4$, 
 we compare the triples $(d_{i-1},\mathbf{s},d_{i+k-1})$ from Theorem~\ref{thm:densities} for $S_{15},Q_{15}$.
For $k=2$ and $k=3$, Tables~\ref{tab:S15d} and~\ref{tab:Q15d} have identical rows 7 and 9, where the triples are ordered for easier comparison as follows.
If needed, we swap $d_{i-1},d_{i+k-1}$ to make sure that the first entry is not larger than the last.
Then we order by the middle bold number $\mathbf s$.
Finally, we lexicographically order the triples with the same middle value $s$.
\medskip

Final rows 11 of Tables~\ref{tab:S15d} and~\ref{tab:Q15d} look different for $k=4$.
More exactly, the rows share three triples (1,{\bf 4},2), (1,{\bf 5},2), (1,{\bf 6},4), but the remaining six triples differ.
However, the density function $\psi_4$ is the \emph{sum} of nine trapezoid functions.
Fig.~\ref{fig:SQ15density4} shows that these sums are equal for $S_{15},Q_{15}$.
Then the sequences $S_{15},Q_{15}$ have identical density functions $\psi_k$ for $k=0,1,2,3,4$, hence for all $k$ by the symmetry and periodicity from Theorem~\ref{thm:symmetries}.
Fig.~\ref{fig:SQ15densities} shows $\psi_k$, $k=0,1,\dots,9$.
\bs
\end{exa}


\begin{figure}[ht]
\includegraphics[width=\textwidth]{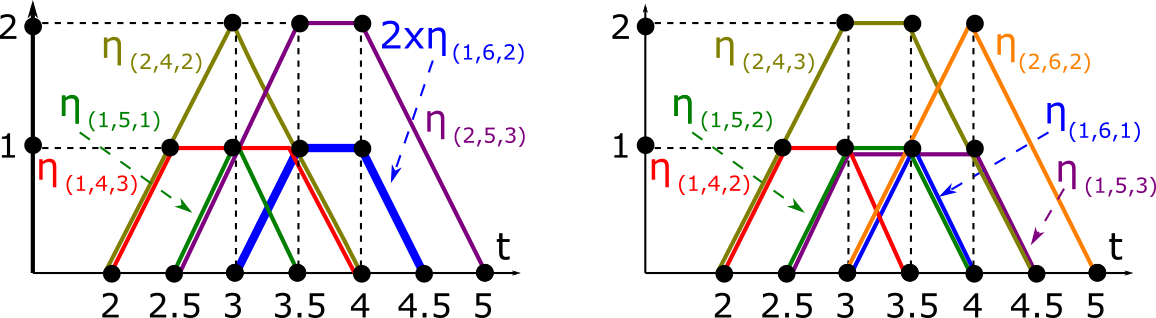}
\caption{The 4th-density function $\psi_4[S_{15}]$ includes the six trapezoid functions on the left, which are replaced by other six trapezoid functions in $\psi_4[Q_{15}]$ on the right, compare the last rows of Tables~\ref{tab:S15d} and~\ref{tab:Q15d}.
However, the sums of these six functions are equal, which can be checked at critical radii: both sums of six functions have $\eta(2.5)=2$, $\eta(3)=5$, $\eta(3.5)=6$, $\eta(4)=4$, $\eta(4.5)=1$.
Hence the periodic sequences $S_{15},Q_{15}$ in Fig.~\ref{fig:SQ15} have identical density functions $\psi_k$ for all $k\geq 0$, see details in Example~\ref{exa:SQ15densities}.} 
\label{fig:SQ15density4}      
\end{figure}

\begin{figure}[h!]
\includegraphics[width=\textwidth]{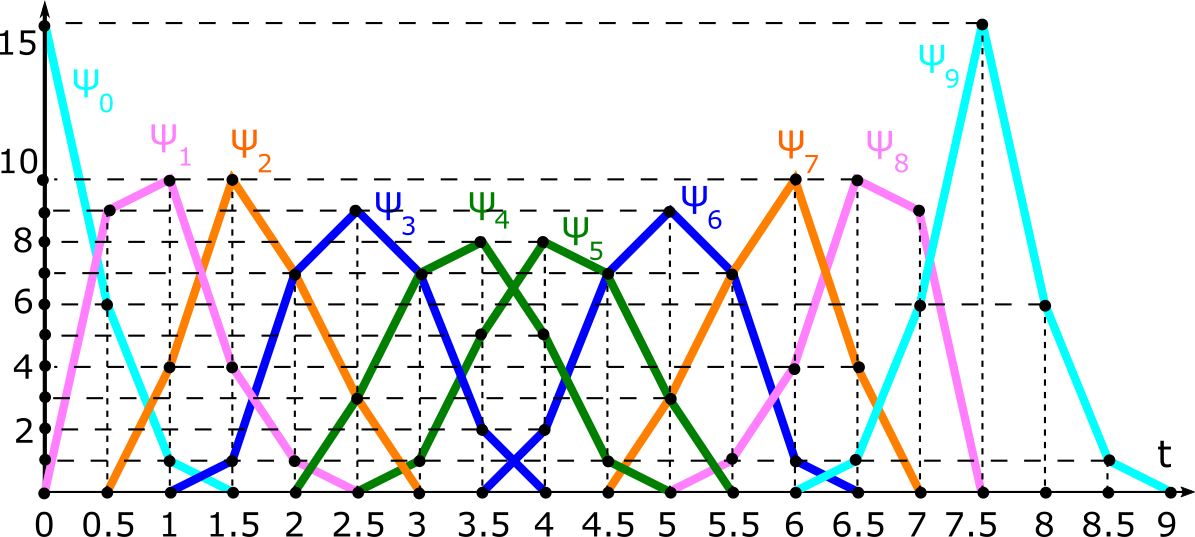}
\caption{The periodic sequences $S_{15},Q_{15}$ in Fig.~\ref{fig:SQ15} have identical density functions $\psi_k(t)$ for all $k\geq 0$.
Both axes are scaled by factor 15.
Theorem~\ref{thm:symmetries} implies
the symmetry $\psi_k(\frac{15}{2}-t)=\psi_{9-k}(t)$, $t\in[0,\frac{15}{2}]$, and periodicity $\psi_9(t+\frac{15}{2})=\psi_0(t)$, $t\geq 0$.}
\label{fig:SQ15densities}      
\end{figure}

Recall that all indices $i$ of distances $d_i$ are considered modulo $m$.  

\begin{cor}[$k$-th density $\rho_k$]
\label{cor:rho-densities}
For any periodic sequence $S=\{p_1,\dots,p_m\}+\Z$ with inter-point distances $d_i=p_{i+1}-p_i$, where $i=1,\dots,m$ and $p_{m+1}=p_1+1$, the $k$-th \emph{density} $\rho_k[S]=\int\limits_{-\infty}^{+\infty} \psi_k(t)dt$ defined as the area under the graph of $\psi_k(t)$ over $\R$ equals
$\rho_k[S]=\dfrac{1}{2}\sum\limits_{i=1}^m d_{i-1} d_{i+k-1}$ for any $k>0$ and $\rho_0[S]=\dfrac{1}{4}\sum\limits_{i=1}^md_i^2$.
\bs
\end{cor}
\begin{proof}
By Theorem~\ref{thm:densities} for $k>0$, each $\psi_k(t)$ is the sum of $m$ trapezoid functions.
Hence $\rho_k[S]$ equals the sum of the areas under the graphs of these trapezoids with corners $(\frac{s}{2},0)$, 
$(\frac{d_{i-1}+s}{2},d)$, 
$(\frac{s+d_{i+k-1}}{2},d)$, 
$(\frac{d_{i-1}+s+d_{i+k-1}}{2},0)$, where 
$d=\min\{d_{i-1},d_{i+k-1}\}$.
The area of each trapezoid is $A_i=\frac{d}{2}(\frac{d_{i-1}+d_{i+k-1}}{2}+|d_{i+k-1}-d_{i-1}|)=\frac{dD}{2}$, where $D=\max\{d_{i-1},d_{i+k-1}\}$.
Then $\rho_k=\sum\limits_{i=1}^m A_i=\frac{1}{2}\sum\limits_{i=1}^md_{i-1}d_{i+k-1}$.
Since $\psi_0(t)=0$ for $t<0$, $\rho_0$ is a half of the area $\rho_m=\dfrac{1}{2}\sum\limits_{i=1}^md_i^2$ under $\psi_m(t)$ due to $\psi_m(\frac{1}{2}\pm t)=\psi_0(t)$ for $t\in[0,\frac{1}{2}]$ by Theorem~\ref{thm:symmetries}, see Fig.~\ref{fig:SQ15densities}.  
\end{proof}

For $S=\{0,\frac{1}{3},\frac{1}{2}\}+\Z$,
Corollary~\ref{cor:rho-densities} gives $\rho_0=\frac{7}{72}$, 
$\rho_1=\rho_2=\frac{11}{12^2}$ as in Fig.~\ref{fig:densities1D}.  

\begin{cor}[generic completeness of $\psi_1$]
\label{cor:gen_complete}
Let $S\subset\R$ be a sequence with period 1 and $m$ points $0\leq p_1<\dots<p_m<1$.
The sequence $S$ is called \emph{generic} if $d_i=p_{i+1}-p_i$ are distinct, where $i=1,\dots,m$ and $p_{m+1}=p_1+1$.
Then any generic $S$ can be reconstructed from the 1st density function $\psi_1[S](t)$ up to isometry in $\R$.
Hence $\psi_1(t)$ is a complete isometry invariant for all generic $S$. 
\bs
\end{cor}
\begin{proof}
As always, one can scale a unit cell of $S$ to the standard interval $[0,1]$ as in Theorem~\ref{thm:densities}.
Hence, up to translation and reflection of $\R$, one can assume that $p_1=0<p_2<1=p_{m+1}$.
It suffices to uniquely locate $p_2,\dots,p_m\in(0,1)$.
\smallskip
 
The 1st density function $\psi_1[S](t)$ is the sum of the trapezoid functions that have the initial gradient $2$ and the corner points
$(0,0)$, 
$(\frac{d_{i-1}}{2},d)$, 
$(\frac{d_{i}}{2},d)$, 
$(\frac{d_{i-1}+d_{i}}{2},0)$, where 
$d=\min\{d_{i-1},d_{i}\}$, $i=1,\dots,m$, all indices are modulo $m$.
\smallskip

Due to a cyclic order of inter-point distances $d_i$, one can assume that the minimum distance is $d_{[1]}=d_1$.
For any $0\leq t\leq d_1$, the function $\psi_1[S](t)$ is linearly increasing with the gradient $2m$.
This gradient drops to $2m-2$ at the first critical radius $t=\frac{d_1}{2}$, which differs from all other larger points $\frac{d_i}{2}$ and $\frac{d_{i-1}+d_i}{2}$ where the gradient of $\psi_1(t)$ changes.
Then the first corner of $\psi_1(t)$ uniquely determines $d_1$ and the second point $p_2=d_1$ of the sequence $S$.
\smallskip

At the radius $t=\frac{d_1}{2}$, subtracting from $\psi_1(t)$ the contribution $(m-1)d_1$ from other still growing $m-1$ trapezoid functions, we get the value $\frac{d_1+d_2}{2}$.
So the first corner of $\psi_1(t)$ also determines the length $d_2=p_3-p_2$ of the second inter-point interval after $[p_1,p_2]$ of the length $d_1$, and the third point $p_3=d_1+d_2$.
\smallskip

Since we know both $d_1,d_2$, we can subtract from $\psi_1(t)$ the whole trapezoid function $\eta(t)$ with the above corners for $i=2$ for all $t\in[0,1]$.
The resulting function $\ti\psi_1(t)$ is the sum of $m-1$ trapezoid functions depending on $m-1$ inter-point distances $d_2,\dots,d_m$. 
We continue analyzing $\ti\psi_1(t)$ by looking at the first corner where its gradient drops from $2m-2$ to $2m-4$, which gives us another pair $(d_{i-1},d_i)$ of successive interval length, and so on.
Since all distances $d_i$ are distinct, the above pairs uniquely determine the ordered sequence $d_1,\dots,d_m$ of all interval lengths, hence the points $p_2,\dots,p_m\in(0,1)$ of the sequence $S$. 
\end{proof}

The recent developments in Periodic Geometry include 
continuous maps of Lattice Isometry Spaces in dimension two \cite{kurlin2022mathematics,bright2021geographic} and three \cite{kurlin2022complete,bright2021welcome}, Pointwise Distance Distributions \cite{widdowson2021pointwise}, and applications to materials science \cite{ropers2021fast,zhu2022analogy}.

%
%
%
\vspace*{-4mm} 
\bibliographystyle{splncs04}
\bibliography{periodic-densities-dimension1}
\end{document}